# Surface corrugations influence on monolayer graphene electromagnetic response

## Yu.A. Firsov[1], N.E. Firsova[2]

[1]A.F.Ioffe Physical-Technical Institute, the Russian Academy of Sciences, St.Petersburg, 194021, Russia,

**email:** yuafirsov@rambler.ru

[2]Institute for Problems of Mechanical Engineering, the Russian Academy of Sciences, St. Petersburg 199178, Bolshoy Prospect V.O. 61, Russia,

[2]St. Petersburg State Polytechnical University, 195231, Politechnicheskaya 29, Russia,

**email:** nef2@mail.ru

**Abstract**

We consider the corrugated monolayer graphene membrane electromagnetic response in terahertz range. We study the generated in irradiated graphene total current (from both valleys) taking into account for the first time both the synthetic electric fields arising due to the (inevitable) presence in graphene of inherent out-of-plane nanodeformations and the double-valleys energy spectrum of Dirac charge particles. Our approach is based on atomistic quantum mechanics used for the description of (1)the valence $\pi - \sigma$ bonds changes generated by activating external periodic electric field and also (2)the mechanism of Dirac electron interaction with this time-dependent perturbation. We consider the problem in the framework of the model of noninteracting Dirac electrons. Assuming surface corrugations not to be very rough we obtain for weak fields the formula for the total current induced in graphene membrane. Our formula describes the curved current lines in the linear in $\vec{E}^{ext}(t)$ approximation for the given graphene surface form. We show that the local direction of current lines is determined by the synthetic electric field whose direction may essentially differ from the one of the external field and depends on the local curvature of the graphene membrane. We also demonstrate that valley currents generated by a linearly external field have nonzero elliptic polarization angles depending on the point $(x, y)$. Valley currents are shown to rotate in opposite directions in different valleys. The results obtained below can be applied to the analysis of different devices in terahertz optics and optoelectronics and the imaging experiments at the Dirac point.

**Introduction**

The successful preparation of one-atomic layer of carbons, i.e. graphene [1-2] gave rise to the development of two-dimensional (2D) physics (see for instance the review [3]). Graphene exhibits a variety of interesting transport phenomena [3-5]. Graphene also shows remarkable optical properties which makes graphene a potential material for nanoelectronics particularly for high-frequency applications. These properties make graphene an ideal photonic and optoelectronic material [6]. One of the most important problems in this field is the investigation of the graphene electromagnetic response. This problem was studied in series of papers using different methods.

Already In 2007 it was understood that graphene's linear electronic dispersion properties should lead to a strongly nonlinear optical behavior at microwave and terahertz frequencies. In [7-8] the kinetic (quasiclassical) transport equation for Dirac quasiparticles to describe first the nonlinear graphene electromagnetic response was obtained. Note that in [7-8] only intraband transitions were taken into account which was reasonable as they considered radiation in terahertz frequency range. For the case of higher optical frequencies Ishikawa [9] used the time-dependent Dirac equation cast into the form of extended optical Bloch equation which allowed him to show that for $\hbar\omega > E_F$ due to the interplay of interband and intraband transitions total nonlinear optical response grows more slowly. The theory of H. Dong et al [10] is consistent with the more precise quantum approach of Ishikawa and gives the same result for $\hbar\omega < E_F$. Interesting new result obtained in [10] is a moving Townes-like spatial solitary waves i.e. soliton (see fig.2 and 3 in [10]) which arises due to inclusion of the term $\nabla f_\mathbf{p}(x, y, t)$ into the quasiclassical transport equation in the form [7-8]..

Most of the previous theoretical studies and experimental observations were devoted to very peculiar properties of the second or third order harmonics excited in graphene by external time-dependant radiation. But there exist situations when one can not confine himself only to these first harmonics for instance for description of strong terahertz emission stimulated from optically pumped graphene etc. Therefore it is necessary to develop general theory of nonlinear electromagnetic response for graphene. In the papers [7-10] mentioned above the first studies of the nonlinear optical response in graphene were published which were not based on the investigation of separate harmonics.

Excellent photophysical properties and large optical nonlinearities with ultrafast response times including fast optical communications promise many potential applications, (see review [11]).

The problem of graphene electromagnetic response was studied also in a series of papers on the basis of pure quantum mechanics using the Floque theory approach (see the review describing this method in [12]). For instance the authors in the framework of this approach were discussing the question whether the dynamical gaps open or not in the quasi-energy spectrum as a result of the interaction of propagating quasiparticles with the external radiation (see [13-14] and others). However the influence of the inevitably existing surface corrugations was not investigated. Note that it would be extremely difficult to take them into account in the framework of the Floque theory approach while on the basis of the self-consistent method suggested in [7-8] it proved to be possible and one can find the corresponding results obtained below.

As it is well known two $\sigma$-bonded interpenetrating triangular sublattices $A$ and $B$ build the graphene honeycomb lattice. The $\pi$ and $\pi^*$ bonds govern the electronic properties near the neutrality point and it is described as electron and hole cones coupled in pairs. Cones of each pair touch each other at the high symmetry point $K$ or $K'$ of the Brillouin zone ($K$ and $K'$ valleys). Near $K$ and $K'$ the energy dispersion is linear if e-e interaction is neglected, and the electronic properties are well described by the Dirac Hamiltonian. Its stationary states are degenerated in the spin and the valley quantum numbers ($\tau = \pm 1$). The last correspond to the pair of the non-equivalent Dirac points in the Brillouin zone. If intervalley scattering is suppressed (which is usually the case because of large distance between these points) this opens new possibilities for applications. This field of activity is called valleytronics.

Many works were devoted to the elucidation of the influence of double-valley spectrum the equilibrium electronic properties of the monolayer graphene and the non-equilibrium time-dependent phenomena in the same material (see for instance [14-15]). The possibility of the separation of the polarized valley currents was discussed in a number of papers (see [16-17]). But in all these papers the role of the inevitably existing inherent ripples in graphene was not taken into account

All the above mentioned methods are based on the one-electron approximation and do not take into consideration manybody effects. The influence of the inherent out-of-plane corrugations and very specific consequences of the double-valley electron spectrum also are not taken into account. To begin with we'll make use of a rather simple approach formulated in [7] and elaborated in [8] for the description of monolayer graphene nonlinear optics and based on the flatland model of noninteracting Dirac 2D electrons. We'll generalize this approach additionally taking into account "internal" time-dependent "synthetic" electric fields induced by radiation influence on the inherent out-of-plane corrugations (ripples, wrinkles etc.). We'll explain below: 1)Why it is important to take into account the out-of-plane corrugations influence the monolayer graphene electromagnetic response and 2)Why it is reasonable to begin the investigation of nonlinear optical response with simplified model of noninteracting 2D Dirac electrons.

1)More than 70 years ago Peierls [18-19] and Landau [20-21] showed that in standard harmonic approximation strictly 2D crystal cannot exist because thermal fluctuations should destroy long range order resulting in "melting" of a lattice at any finite temperature. Actually (see [22]) all the observed monolayer graphene samples have inherent stable out-of-plane deformations (ripples, bubbles, wrinkles etc.). These corrugations lead to the arising of pseudomagnetic fields (gauge fields), (see for instance review [23]).These inevitably existing fields in graphene are about several Tesla. It is interesting that such fields not long ago were artificially created in another nontrivial system – in rubudium Bose-Einstein condensate (BEC). This pseudomagnetic field was created as time-dependent via interaction of BEC atoms with laser light. This led to the appearance of the so-called synthetic electric fields [24].

Time dependent (due to the presence of flexural phonons (f-ph)) gauge field and its time-derivative, i.e. synthetic electric fields for monolayer graphene, was first considered in [25] where it was shown that (f-ph) damping rate is determined by interaction with these fields.

In graphene nanoresonators vector potential depends on time due to the influence of the external time-dependent electromotive force. It was shown in [26] that this generates synthetic electric fields and associated with them currents. It was found in [26] that this leads to a new mechanism of the dissipation ("heating") and determines the loss which strongly influences the quality factor (about 30%).

Below to begin with we are going to consider the graphene corrugations influence the linear electromagnetic response. If in paper [26] considered frequencies were not higher than 1GHz in the present investigation we analyze the corrugations influence monolayer graphene response to electromagnetic radiation of terahertz [300GHz-3THz] and infrared [201THz-790THz] frequency range.

2)There are many cases (see for instance the review [27] and discussion below), when in the 2D system of interacting Dirac electrons many-body effects should be taken into account. However sometimes this interaction does not play significant role (see below). It was shown that together with quasiparticles ("massless electrons" and

"holes") an important role may play plasmons (collective charge density excitations) and "plasmarons" [28] ("dressed" quasiparticle exitations coupled to plasmons). These predictions are based on a very specific form of quasiparticle spectral function obtained theoretically [29-30] and observed experimentally by variety of direct and indirect methods: electron energy-loss spectroscopy (EELS),(see in [31]), angle-resolved photoemission spectroscopy (ARPES) [28], scanning tunneling spectroscopy (STS) [32] on exfoliated graphene sheets and on epitaxial graphene samples.

Large efforts were made to observe these predictions in optical measurements. As a rule, indirect methods are used "by engineering plasmons coupling to infrared light in a number of intriguing ways", [27], by making more strong light-matter interaction and alleviating the momentum mismatch between plasmon modes and the incident radiation. To illustrate this idea we give as an example Mikhailov formula for the plasmon enhanced second harmonic generation in graphene [33]

$$I_{2q,2\omega}^{graph} \sim \left(I_{q,\omega}^{ext}\right)^2 / \left\{ \left[ \left(\omega^2 - \frac{\omega_p^2(q)}{2}\right)^2 + \frac{\omega^2 \gamma^2}{4} \right] \left[ \left(\omega^2 - \omega_p^2(q)\right)^2 + \omega^2 \gamma^2 \right]^2 \right\}$$

Here $I_{q,\omega}^{ext}$ is the intensity of the incident light, $I_{2q,2\omega}^{graph}$ is the intensity of the second-harmonic wave in graphene, $\omega_p(q) = \sqrt{\frac{8E_F}{h\varepsilon} \frac{\pi}{2} \frac{e^2}{h} q}$ is the plasmon frequency, $q$ is the Plasmon wave vector, $\gamma$ is the momentum scattering rate for plasmon. Denominator in the last formula strongly decreases as $\omega \to \omega_p(q)$ and $I_{2q,2\omega}^{graph}$ strongly increases. However one should not forget about the momentum conservation law: $q_{light} = q_{plasm}$. Let us estimate the ratio $q_{pl}/q_{light}$ for $\omega = \omega_{light} \approx \omega_{pl}$. As $\omega_{light} = cq_{light}$ for external field we obtain $q_{pl}/q_{light} = \frac{h\omega_{light}}{E_F} \left(\frac{e^2}{hc}\right)^{-1}$ which shows that it is very difficult to satisfy simultaneously the both conditions $\omega_{light} \approx \omega_{pl}$ and $q_{light} \approx q_{pl}$. However in 2011 Z.Fei, D.N.Basov et al [34] managed to overcome these difficulties creating stronger light-matter interaction by confining mid-IR radiation at the apex of a nanoscale tip of atomic force microscopy. This turned to be a very effective method to alleviate the momentum mismatch between plasmon modes and incident radiation. Through infrared nanoimaging the authors of [34] have shown that graphene/$SiO_2$/Si back-gated structures may support a propagating of surface plasmons.

Independently in Chen, J.et al [35] it was made similar discovery using graphene films deposited by a gas rather than peeled from graphite. They used near field produced by a permanent dipole. As far as we know, up to now it was not shown experimentally that homogeneous light radiation of a single layer graphene membrane may excite plasmons.

It is interesting to note that L.Crassee et al [36] have shown that in graphene epitaxially grown on SiC the Drude absorption is transformed into a strong terahertz plasmon peak due to natural nanoscale substrate terraces and wrinkles. This is a natural confinement potential and it does not require special lithographie patterning.

Thus we come to conclusion that under incident homogenous radiation surface plasmons (SP) cannot be driven without using special efforts: either by using AFM tip to confine the incident radiation to $q$ nanoscale region around the tip as Z.Fei, D.N.Basov et al [34] or by exciting SP in graphene micro-ribbon array or by using stacked graphene microdiscs to realize electromagnetic radiation shield with 97,5% effectiveness [27].

Measurement of the linear optical conductivity of graphene by Mak et al [37] do not contradict to the theoretical results predicted within a model of non-interacting massless Dirac fermions. Taking into account e-e interactions (see [38]) with the renormalization group techniques results to the notable modifications but do not reveal plasmon excitation without applying any special efforts (see above). At present we are not aware of any experimental investigations of graphene nonlinear electromagnetic response to homogenous radiation in which SP were displayed . Theoretical investigations of the single layer graphene transmittance (T) in nonlinear case shows very small (~2%) changes of T (see [39]). Therefore we shall use the model of noninteracting Dirac electrons (especially for the graphene membrane placed on the Irridium111 substrate due to the effective screening, [40], this theoretical assumption is realistic) as it was done in [7-8] to study electromagnetic response of the monolayer graphene membrane in terahertz range.

However while in [7-8] they considered the flat model we investigate the graphene sheet with corrugations. As far as the out-of-plane nanodeformations inevitably exist as it follows from Landau-Peierls theorem [18-21] it is very important to study these corrugations influence the electromagnetic response which we investigate below. For the first time the corrugations influence in nonstationary case was taken into account in [26] where we proposed the method of treating this problem. So we consider the problem on the basis of the kinetic (quasiclassical) transport equation with time-dependent synthetic electric fields generated by corrugations and the double-valley structure of the electronic spectrum taken into account. We find in linear approximation the formulas describing the curved current lines as an electromagnetic response while in flat case they were straight ([7-8]). The curvature of the current lines proved to be correlated with the form of the corrugated membrane surface. We also study valley currents elliptic polarization angles in irradiated graphene as functions of a point (x,y) for given surface form.

## The total current induced in irradiated corrugated graphene membrane in linear approximation

For the flat graphene membrane this problem was studied in [7-8]. Let us consider electromagnetic response for monolayer graphene membrane with out-of-plane corrugations described by the function $z = h(x, y)$. This curvature leads to the arising of the so called gauge field $\vec{A}(x, y)$ (see review [41]). The formulae for them taking into consideration the nearest neighbor hopping (see review [41]) and the formulae for them were obtained in [42] (see also [43-44])

$$H = \int d^2r \, \psi^*(\vec{r}) \left[ \vec{\sigma} \cdot \left( i\hbar v_F \nabla + e\vec{A} \right) - \mu \right] \psi(\vec{r}),$$

$$A_x(\vec{r}) + iA_y(\vec{r}) = -\sum_j \delta t_j(\vec{r}) e^{i\mathbf{u}_j \mathbf{K}} = -\frac{\epsilon_1}{2} \sum_j \left[ (\mathbf{u}_j \cdot \nabla) \nabla h \right]^2 e^{i\mathbf{u}_j \mathbf{K}} \quad (1)$$

$$A_x = -\frac{1}{2} A^0 \left[ (h_{xx})^2 - (h_{yy})^2 \right] a^2, \quad A_y = A^0 \left[ h_{xy}(h_{xx} + h_{yy}) \right] a^2 \quad (2)$$

$$A^0 = 3/4 \cdot \epsilon_1/e \cdot c/v_F \quad (3)$$

Here $\epsilon_1 = 2{,}89 ev$, $\vec{K} = a^{-1}(4\pi/3\sqrt{3}, 0)$ corresponds to a Dirac point and $t_j$ - exchange integral with the $j$-th nearest neighbor, $j = 1,2,3$, and $A^0$ has the same dimensionality as vector potential. In formulae for $A_x, A_y$ in (2) the products of the expressions in square brackets multiplied by $a^2$ are dimensionless, i.e. they are numerical coefficients. Their magnitudes are functions of a point (x,y) determined by the form $z = h(x,y)$ of nanodeformations (such surface corrugations as ripples, wrinkles etc.) and also by the lattice constant $a$ value for the current moment of time. Besides, taking into account the hopping between next nearest neighbors in [42] it was found that the curvature generates the scalar potential $\Phi(\vec{r})$

$$\mathcal{H} = H + \delta H, \quad \delta H = e \int d^2 r\, \psi^*(\vec{r}) \Phi(\vec{r}) \psi(\vec{r}) \tag{4}$$

$$e\Phi(\vec{r}) = -3 \sum_{\vec{v}} \delta t'_i(\vec{v}) e^{-i\vec{v} Q_1} \approx -\frac{3}{4} \alpha a^2 [\nabla^2 h(\vec{r})]^2 \tag{5}$$

where $\alpha = 9.23 eV$ is an energy scale associated with the mixing between $\pi - \sigma$ bonds.

Under the action of the external electromagnetic field the existence of the inherent out-of-plane deformations (bubbles, ripples…) leads to the arising of quantum effects due to the changing of overlap for different valence bonds having quantum origin. The vector-potential $\vec{A}(x,y,t)$ and scalar potential $\Phi(x,y,t)$ begin to depend on time which leads to the arising of so called synthetic electric field

$$\vec{E}^{syn}(x,y,t) = \vec{E}_1^{syn} + \vec{E}_2^{syn} \tag{6}$$

The first term in (6) was calculated in [26] for the case $\vec{E}^{ext}(t) = \vec{E}_0 \sin \omega t$

$$\vec{E}_1^{syn} = -c^{-1} \partial \vec{A}(x,y,t)/\partial t \tag{7}$$

We assume for simplicity that the membrane surface does not change and consider the linear approximation $\Delta a(t) = \eta E_0 \sin \omega t$. Making use of (1)-(3) we find for external actuating periodic electric field $\vec{E}^{ext}(t) = \vec{E}_0 \sin \omega t$ the formulae for $(\vec{E}_1^{syn})_x$, $(\vec{E}_1^{syn})_y$ in the following form

$$(\vec{E}_1^{syn})_x (x,y,t) = -E^0(\omega) E_{0x} I_x(x,y), \cos \omega t$$

$$(\vec{E}_1^{syn})_y (x,y,t) = -E^0(\omega) E_{0y} I_y (x,y) \cos \omega t$$

where

$$E^0(\omega) = 3/4 \cdot \epsilon_1/e \cdot \omega/v_F, \quad I_x = a\eta(h_{xx}^2 - h_{yy}^2), \quad I_y = 2a\eta(h_{xx} + h_{yy}) h_{xy}, \tag{8}$$

We would like to pay the reader's attention to the fact that $E^{syn} \sim \omega$. This means that "synthetic" electric fields (in contrast to "pseudomagnetic" fields) are of a dynamic origin. Coefficient $\eta$ determines time corrections to the lattice constant $\Delta a(t)$ i.e. $a(t) = a_0 + \Delta a(t)$ and has the dimension $cm^2 V^{-1}$. For not very high $\omega$ coefficient $\eta$ can be measured experimentally using field effect and applying scanning tunneling microscopy (STM) and spectroscopy methods which were used for the imaging of curvature in graphene. Note that quantities $I_x, I_y$ do not turn to zero because of the presence in graphene of such deformations as ripples, wrinkles and so on. The second term in (6) is a sum of the stationary field $\nabla \Phi(x, y, 0)$ and the nonstationary term which is small compared to $\vec{E}_1^{syn}$ provided the surface corrugations are not too rough (which we assume). So we neglect the second term in (6). Note that inequality $|\vec{E}_2^{syn}| \ll |\vec{E}_1^{syn}|$ is due to the fact that $\vec{E}_1^{syn}$ was generated by the gauge field $\vec{A}$ which takes into account the nearest neighbors interaction while $\vec{E}_2^{syn}$ was generated by the scalar potential $\Phi$ which arises due to the next to nearest neighbors interaction. So we have

$$\vec{E}_x^{syn}(x, y, t) = -E_{0x}^{syn} \cos \omega t = -E_{0x} D^{(x)}(x, y) \cos \omega t, \quad D^{(x)}(x, y) = E^0(\omega) I_x(x, y) \quad (9),$$

$$\vec{E}_y^{syn}(x, y, t) = -E_{0y}^{syn} \cos \omega t = -E_{0y} D^{(y)}(x, y) \cos \omega t, \quad D^{(y)}(x, y) = E^0(\omega) I_y(x, y). \quad (10)$$

Our approach to obtain the formula for the total current in both valleys taking into account the curvature is to co$\cos \omega t$nsider the kinetic equation written in the standard form

$$\frac{\partial f(\vec{p}, \vec{r}, t)}{\partial t} = -e\big[\vec{E}^{ext}(t) + \vec{E}^{syn}(\vec{r}, t)\big]\frac{\partial f(\vec{p}, \vec{r}, t)}{\partial \vec{p}} - \vec{v}\frac{\partial f(\vec{p}, \vec{r}, t)}{\partial \vec{r}} - \frac{f(\vec{p}, \vec{r}, t) - f_0(\vec{p}, \vec{r})}{\tau}$$

Here $\vec{E}^{ext}(t) = \vec{E}_0 \sin \omega t$ is a external periodic field (we assume that it does not depend on $x, y$ as we consider incident field wavelength to be much greater than membrane sizes); $\vec{E}^{syn}(\vec{r}, t)$ (see (6)-(10) is the synthetic electric field generated by the corrugations of the graphene membrane excited by the external periodic electric field $\vec{E}^{ext}(t)$. We shall see later that the last term in kinetic equation can be neglected i.e. we ignore collisions of electrons with impurities, phonons and other lattice imperfections as by high frequencies more important proved to be another time scale related with the radiative decay rate which guarantees the more rapid loss than $\tau$. The radiative decay rate was calculated in [8] in the flat case (see (35) in [8]). We neglect also the gradient term in kinetic equation as it is small provided the surface corrugations are not very rough which we assume. So we consider the kinetic equation in the form

$$\frac{\partial f(\vec{p}, \vec{r}, t)}{\partial t} - e\big[\vec{E}^{ext}(t) + \vec{E}^{syn}(\vec{r}, t)\big]\frac{\partial f(\vec{p}, \vec{r}, t)}{\partial \vec{p}} = 0 \tag{11}$$

The equation (11) for the temperature close to zero has the exact solutions

$$f(\vec{p}, \vec{r}, t) = F_0(|\vec{p} - \vec{p}_0(\vec{r}, t)|), \qquad F_0(p) = \left[1 + exp\left(\frac{vp - \mu}{kT}\right)\right]^{-1} \tag{12}$$

where $F_0(p)$ is the Fermi-Dirac distribution function, and

$$\vec{p}_0(\vec{r}, t) = -e \int_{-\infty}^{t} [\vec{E}^{ext}(t') + \vec{E}^{syn}(\vec{r}, t')] dt'$$

Is the solution of the single particle classical equation of motion. The electrical current then takes the form

$$\vec{j}(\vec{r}, t) = -\frac{g_s e^2 v_F}{(2\pi\hbar)^2} \iint dp_x dp_y \frac{\vec{p}}{\sqrt{p_x^2 + p_x^2}} F_0(|\vec{p} - \vec{p}_0(\vec{r}, t)|) \tag{13}$$

Comments about formula (13) see in [45]. If the temperature is zero, $T = 0$, and the chemical potential is finite, $\mu > 0$, the current $\vec{j}(\vec{r}, t)$ can be presented in the form

$$j_x = \frac{g_s e v_F}{(2\pi\hbar)^2} p_F^2 G_x(Q_x, Q_y) \frac{2P_{0x}}{\sqrt{1 + P_{0x}^2 + P_{0y}^2}}, j_y = \frac{g_s e v_F}{(2\pi\hbar)^2} p_F^2 G_y(Q_x, Q_y) \frac{2P_{0y}}{\sqrt{1 + P_{0x}^2 + P_{0y}^2}} \tag{14}$$

where

$$G_x(Q_x, Q_y) = \frac{1}{Q_x} \int_0^{\pi/2} \left[\sqrt{1 + Q_y \sin \varphi + Q_x \cos \varphi} - \sqrt{1 + Q_y \sin \varphi - Q_x \cos \varphi}\right] \cos \varphi \, d\varphi$$

$$G_y(Q_x, Q_y) = \frac{1}{Q_y} \int_0^{\pi/2} \left[\sqrt{1 + Q_x \cos \varphi + Q_y \sin \varphi} - \sqrt{1 + Q_x \cos \varphi - Q_y \sin \varphi}\right] \sin \varphi \, d\varphi$$

$$Q_{x,y} = \frac{2P_{0x,0y}}{1 + P_{0x}^2 + P_{0y}^2}, \qquad P_{ox,oy} = \frac{\vec{p}_{ox,0y}}{p_F} \tag{15}$$

The equation (11) and its solution (12) is the generalization of (12)-(14) in [8] where the flat case was considered without taking into account the surface corrugations. Note that according to the Landau-Peierls theorem demonstrated in [18-21] these corrugations exist always and consequently our generalization is essential. If the external field $\vec{E}^{ext}(t)$ is small (and consequently $\vec{E}^{syn}(\vec{r}, t)$ is small) and $P_{0x,0y} \ll 1$, and $G_{x,y}(Q_x, Q_y) \approx 1$, we obtain

$$\vec{j}_x \approx -\frac{n_s e^2 v_F}{p_F} \int_{-\infty}^{t} [\vec{E}_x^{ext} + \vec{E}_x^{syn}] dt' \tag{16}$$

$$\vec{j}_y \approx -\frac{n_s e^2 v_F}{p_F} \int_{-\infty}^{t} [\vec{E}_y^{ext} + \vec{E}_y^{syn}] dt' \tag{17}$$

$$n_s \equiv \frac{g_s p_F^2}{4\pi\hbar^2} = \frac{g_s \mu^2}{4\pi\hbar^2 v_F^2}. \tag{18}$$

Here $n_s$ is the density of electrons in the upper band, $g_s = 2$. We can represent these formulas in the form

$$\vec{J}_x \approx \sigma_0 E_{0x}[\cos\omega t + D^{(x)}(x,y)\sin\omega t], \quad \vec{J}_y \approx \sigma_0 E_{0y}[\cos\omega t + D^{(y)}(x,y)\sin\omega t] \quad (19)$$

$$\sigma_0 = \frac{n_s e^2 v_F}{p_F \omega}, \quad (20)$$

$$D^{(x)}(x,y) = E^0(\omega) I_x(x,y), \quad D^{(y)}(x,y) = E^0(\omega) I_y(x,y) \quad (21)$$

These formulas for the current density in the linear approximation describe distortions of the current line, its dependence on the point $(x,y)$. It means that the graphene membrane curvature leads to the curving of the current lines, the total current through the section conserving the constant value. So there is an important difference from the usual percolation description where the origin of the curved current lines is due to the random potential. Comments about the formula (20) see in [46].

We found in (16)-(17) the current density $\vec{j}^\pm$ in one of the vallies $K^+$ or $K^-$ (usually they denote them $K$ and $K'$ but it is less convenient). Let us consider the total current density in $K^+$ and $K^-$

$$\vec{j}^{tot} = \vec{j}^+ + \vec{j}^- \quad (22)$$

It was found (see (12) in [47]) that for any $m > 0$ i.e. when there is a gap in spectrum we have

$$\vec{j}^{tot} \sim 2\frac{m}{|m|}\vec{E}^{syn} \quad (23)$$

(in [47] $\vec{E}^{syn}$ is denoted as $\vec{E}^{el}$). This fact follows from the time-reversal symmetry for flat graphene which says that the current densities in different valleys have equal values but opposite directions. So for $m \to 0$ we have (for more details see below)

$$\vec{j}_x^{tot} \approx -2\frac{n_s e^2 V}{p_F}\int_{-\infty}^{t} \vec{E}_x^{syn} dt' \quad (24)$$

$$\vec{j}_y^{tot} \approx -2\frac{n_s e^2 V}{p_F}\int_{-\infty}^{t} \vec{E}_y^{syn} dt' \quad (25)$$

These formulas mean that we have $\vec{j}^{tot} \neq 0$ i.e. the violation of the time-reversal symmetry is due to the presence of the surface curvature. Using (9)-(10) we can rewrite (24), (25) as follows

$$\vec{j}_x^{tot}(x,y) = \sigma_{xx}(x,y) E_{0x}\sin\omega t, \quad \vec{j}_y^{tot}(x,y) = \sigma_{yy}(x,y) E_{0y}\sin\omega t \quad (26)$$

$$\sigma_{xx}(x,y) = 2\sigma_0 D^{(x)}(x,y), \quad \sigma_{yy}(x,y) = 2\sigma_0 D^{(y)}(x,y). \quad (27)$$

From the obtained formulae we see that the total current $\vec{j}^{tot}$ (26) induced by the field $\vec{E}^{ext}(t) = \vec{E}_0 \sin\omega t$ either is proportional to the incident field with the conductivity tensor (27) depending on the point $(x,y)$ or is proportional to the synthetic field

$\vec{E}^{syn}(x,y,t)$ and the conductivity does not depend on the point and is equal to $\sigma_0$. From the formulas (9)-(10) it is clear that the really acting synthetic field is not directed along $\vec{E}^{ext}$. Because of the double-valley energy spectrum the acting field is equal to $2\vec{E}^{syn}(x,y)$. Therefore the local current proportional to $\vec{E}^{syn}(x,y)$ may flow along a very complicated trajectory. This is a very curious property of the "breathing" corrugated surface which we do not meet in any other models of disordered systems.

Formulas (26),(27) show that if surface is corrugated (which is always true) the total current is a curved line which depends on the form $z = h(x,y)$ of the surface while in flatland we have straight current lines in each valley and because of the time reversal symmetry the total current of both valleys is equal to zero. We also see that the sign of current components $j_x^{tot}$ and $j_y^{tot}$ in the point $(x,y)$ depends on the sign of $D^{(x)}(x,y)$ and $D^{(y)}(x,y)$ which is determined by the form of the curvature $z = h(x,y)$ in the point $(x,y)$.

The formulae (26), (27) concretize the formula (23) for topological currents considered in [47]. In our concretized formulae the induced total current depends on the form $z = h(x,y)$ of the graphene surface.

### Graphene surface corrugations and valley currents

Last years in a series of papers the problem of separation from the total current the one of the two valley currents was discussed. For instance theoretical ideas how to organize experiments isolating one of two valley currents we can find in [48], [49]. If it were a success it would give new possibilities in valleytronics and new applications would become available. So it would be interesting to analyze valley currents considered in our problem. Below we shall show that in graphene irradiated by linear polarized external electric field $\vec{E}^{ext}(t) = \vec{E}_0 \sin \omega t$ the elliptically polarized valley currents appear. We shall use the formulae for valley currents obtained above to analyze how the acquired angle of ellipticity depends on the form of the graphene surface corrugations.

Note that using Floquet method in flatland and one-valley model it was predicted in [50] that a laser field can produce observable "dynamical" (existing only during the radiation process) bandgap in the electronic spectrum of graphene. It was shown there also that the width of the gap strongly depends on the incident light polarization angle. Namely for the linear polarization of $\vec{E}^{ext}(t)$ there is no gap and for the elliptical one the more is elliptic polarization angle the broader is the gap reaching its maximum for the circular polarization (see Fig.1 in [50]).

To compare with our results let us transform the obtained in (19)-(21) formulae for the induced valley current density in one valley for given form of the membrane surface as follows

$$j_x(x,y) = \sigma_0 E_{0x}\sqrt{1 + \left(D^{(x)}(x,y)\right)^2}\,\cos\left[\omega t - \delta^{(x)}(x,y)\right] \tag{28}$$

$$j_y(x,y) = \sigma_0 E_{0y}\sqrt{1 + \left(D^{(y)}(x,y)\right)^2}\,\cos\left[\omega t - \delta^{(y)}(x,y)\right] \tag{29}$$

In (28), (29) the phases $\delta^{(x)}(x,y)$, $\delta^{(y)}(x,y)$ are defined as follows

$$\sin\delta^{(x)}(x,y) = \frac{D^{(x)}(x,y)}{\sqrt{1 + \left(D^{(x)}(x,y)\right)^2}}, \quad \sin\delta^{(y)}(x,y) = \frac{D^{(y)}(x,y)}{\sqrt{1 + \left(D^{(y)}(x,y)\right)^2}}, \tag{30}$$

Let us introduce also

$$\sin\delta(x,y) = \sin\left[\delta^{(x)}(x,y) - \delta^{(y)}(x,y)\right] = \frac{D(x,y)}{\sqrt{1 + \left(D^{(x)}(x,y)\right)^2}\sqrt{1 + \left(D^{(y)}(x,y)\right)^2}} \tag{31}$$

where

$$D(x,y) = D^{(x)}(x,y) - D^{(y)}(x,y) = E^0(\omega)\eta\left[I_x(x,y) - I_y(x,y)\right] \approx$$

$$\approx E^0(\omega)\eta a\left(h_{xx}(x,y) + h_{yy}(x,y)\right)\left[\left(h_{xx}(x,y) - h_{yy}(x,y)\right) - 2h_{xy}(x,y)\right] \tag{32}$$

Note that for the "second" valley we get the function (32) with the opposite sign (instead of $\delta^{(x,y)}(x,y)$ we have $\pi - \delta^{\prime(x,y)}(x,y)$). We see that the formulae (28)-(32) describe elliptically polarized valley currents which circulate clockwise if $D(x,y) \geq 0$ and in the opposite direction if $D(x,y) \leq 0$. Here from we get that valley currents in different valleys in the same point rotate in opposite directions while their sum i.e. the total current is linearly polarized in the direction of $\vec{E}^{syn}(x,y,t)$.

We showed that the incident linearly polarized field will be inevitably transformed (due to corrugations) into the elliptically polarized valley currents. So our results mean that for the linearly polarized incident field we shall have a narrow dynamic gap in graphene electronic spectrum. So our irradiated by linearly polarized light graphene membrane due to corrugations has dynamic gap in spectrum while in flatland model it is not possible, [50].

## Conclusions

In this paper we obtained and analyzed the expressions for the total current which is a monolayer graphene membrane electromagnetic linear response to an external time-dependent weak electric field $\vec{E}^{ext}(t) = \vec{E}_0 \sin\omega t$. To solve the problem we took into account quantum "synthetic" electric fields arising due to the inevitably existing out of plane nanodeformations "breathing" under the influence of external time-dependent field $E^{ext}(t)$. We used the formulae for them obtained in [26] where these synthetic fields were introduced. We also took into consideration a very special feature i.e. two valleys coupling in the graphene energy spectrum in the first Brillouen band. Up to now both of these factors were not taken into account jointly in the considered problem.

The equation (11) we used to analyze our problem is in fact the equation of motion of Dirac electrons (taking into account their distribution according to Fermi-Dirac statistics) in nonstationary external and synthetic electric fields but without taking into consideration e-e interaction (possibility of such an approach was at length discussed in the Introduction). As a matter of fact as a result of radiation in graphene "breathing metrics" appears which leads to the arising of $\vec{j}^{tot}$ as an electromagnetic response. So actually we get mathematical problem with nonstationary differential equation on the curved surface with metrics periodically depending on time. We analysed this problem after corresponding simplifications. So as it was mentioned above the nature of the problem studied here differs from the usually considered percolation model.

We obtained the expressions (19)-(21) for electromagnetic response valley current densities $j_x(x,y)$, $j_y(x,y)$ in linear approximation (i.e. for weak external field). Note that phenomenological constant $\eta$ (see (8), (19)-(21)) can be studied experimentally using STM and STS methods.

The formulae we got for the induced total current density concretize the results obtained in [47]. Our equations describe the total current relation with the form of the membrane surface curvature which is interesting from theoretical point of view. The found formulae demonstrate that in the arbitrary point $(x,y)$ current lines are essentially curved due to the influence of local "synthetic" electric fields stipulated by changing in time of the curvature (while in flatland the current lines are straight). Note that if we introduce the so-called "total conductivity" $\sigma^{tot} = \sqrt{\sigma_{xx}^2 + \sigma_{yy}^2}$ then using (8)-(10), (27) we find

$$\sigma^{tot} = 2a\eta E_0(\omega)\sigma_0 \ |h_{xx} + h_{yy}| \sqrt{[(h_{xx} - h_{yy})^2 + 4h_{xy}^2]} \tag{33}$$

This formula shows that in points with zero curvature $h_{xx} + h_{yy} = 0$ the total conductivity is equal to zero. For instance we have the zero total conductivity on the boundary of the bubble which is directed up or down. It is interesting that in [51] in the Fig1, middle colored plot where conductivity was measured we see light yellow lines where this conductivity has minimal values. May be they correlate with the lines of zero curvature which we got.

We found that the first electromagnetic response i.e. the currents generated in graphene membrane due to the time-dependent gauge fields $\vec{A}$ is much stronger that the second contribution to the electromagnetic response due to the time-dependent scalar potential Φ. The reason is that the gauge field $\vec{A}$ takes into account only the nearest neighbors interaction while the scalar potential Φ is generated by the next to nearest neighbors interactions.

Comparing our results with those obtained in [50] on the basis of Floquet theory for one-valley model neglecting the surface corrugations influence we see that in [50] they say that a dynamic gap in the irradiated graphene electronic spectrum can appear only if the polarization angle in the incident radiation is not equal to zero while we get the possibility of a gap also in case of the linear polarization (note that in [47] it was assumed the gap in spectrum). Also it was shown in [52] for flatland model that because of the interaction between pseudospin and the pseudomagnetic field the valley degeneracy may be lifted and a gap may appear. However in real membrane which is always curved a bandgap in static case may be "washed away" by disorder due to corrugations.

From the obtained formulas it is seen that the rougher is the curvature of the surface in the point the stronger the current line is curved having a tendency to convert into closed ellipse for large values of the curvature and the frequency of the external field.

The direction the valley current line turns to can be clockwise or the opposite one depending on the sign of $D(x, y)$ which is determined by the form $z = h(x, y)$ of the membrane surface. This circumstance can be interpreted assuming that the sign of $D(x, y)$ is also related with the surplus of the electrons or holes in the vicinity of the considered point which curves the current lines to the right or to the left correspondingly. We found that valley currents in different valleys rotate in opposite directions. This result could be of use for the problem of valleytronics to separate a valley current from certain valley. If such a separation were a success it would give new possibilities for interesting applications.

Earlier in [26] we found that the contribution of surface corrugations into the value of the quality factor of the graphene nanoresonator can make it 30% less. So we see that in constructing graphene devices it is very important to take into account the surface corrugations influence. Note that in [53] it was mentioned the significant $Q$-factor increasing when reduction in out-of-plane buckling heights by using tensile strain. This strain-engineering method to reduce loss in nanoresonators was suggested for the first time in [54] as it followed there from theoretical arguments. Note also that the prediction of the quality factor increase due to the magnetic field switching on perpendicular to the graphene membrane was published at first in [54] while the experimental investigation of the dependence of $Q$ on $B_\perp$ was done in [55] (see Fig3(c)) in quantum limit for the temperature $T = 5K$. The increase of $Q$ in quantizing magnetic field $B_\perp$ was really observed and its value proved to be about 30%. So there is experimental demonstration of the correctness of our approach to the estimation of surface corrugation influence the nonstationary processes in graphene. The results obtained in the present paper could be also useful in this respect. They can be applied as well to the analysis of different devices in terahertz optics, optoelectronics and in the imaging experiments at the Dirac point.

In this paper we did not consider the radiation loss in the electromagnetic response. This processes will be studied on the basis of the nonlinear selfconsistent equations which we are planning to describe and to analyze in the next publication. Also we plan to study the strong external fields influence graphene which would lead for instance to the essential changing of the sinusoidal form of the induced current due to the nonlinearity. This question will be investigated in the next paper as well.

**Referencies**

done in [43] one gets (2), (3) from the generalized formulae [43]. We consider for simplicity this case and hence use the formulae (2), (3)

[45]According to (13) the system of Dirac electrons automatically regulates the behavior of $p_0(t)$ and does not warm up because the energy absorbed from electromagnetic field is radiated. This process will be described by generalization to the case of curved surface the self-consistent approximation suggested in [7-8] for flatland in order to take into account a feed back from the exited surface currents.

[46]Actually eq. (20) will be further improved by including the radiative decay rate. For this purpose, the self-consistent approximation considered in [7,8] for flatland in order to take into account feed back from the excited surface currents will be generalized to the curved surface case. Note that the described procedure will not change functions $D^{(x,y)}(x,y)$ related to corrugations influence.